# Some drastic improvements found in the analysis of routing protocol for the Bluetooth technology using scatternet

**Yusuf Perwej**
M.Tech , MCA
Computer Science & Information System
Jazan University, KSA
yusufperwej@gmail.com

**Dr.Kashiful Haq**
P. hd
IT department, College of Applied Sciences
Ibri, Sultanate of Oman
kashifulhaq@rediffmail.com

**Uruj Jaleel**
M.Tech , MCA
Computer Science & Information System
Jazan University, KSA
urujjaleel@yahoo.co.in

**Sharad Saxena**
M.Tech , MCA
Shobhit University
Meerut , UP ( India )
sharadzoom@gmail.com

**ABSTRACT**

Bluetooth is a promising wireless technology that enables portable devices to form short-range wireless ad hoc networks. Unlike wireless LAN, the communication of Bluetooth devices follow a strict master slave relationship, that is, it is not possible for a slave device to directly communicate with another slave device even though they are within the radio coverage of each other. For inter piconet communication, a scatternet has to be formed, in which some Bluetooth devices have to act as bridge nodes between piconets. The Scatternet formed have following properties in which they are connected i.e every Bluetooth device can be reached from every other device, Piconet size is limited to eight nodes [1]. The author of this research paper have studied different type of routing protocol and have made efforts to improve throughput and reduce packet loss due to failure in the routing loop and increased mobility and improve the cohesive network structure, resolve the change topology conflicts [2], and a successful & efficient transfer of packet from source to destination.

**Keywords:** Bluetooth, Scatternet, Piconet, Ad hoc, TDD (Time Division Duplex), PAN (Personal Area Network), master, slave, routing protocol.

## 1  INTRODUCTION

Bluetooth is a low-cost, low-power wireless networking technology, mainly intended be a cable Replacement between portable and /or fixed electronic devices. The true potential of the technology begins to reveal itself when it supports low-cost solutions to creating ad-hoc mobile personal area networks. The idea was born in 1994 when Ericsson Mobile Communications decided to investigate the feasibility of a cheap radio interface between mobile phones and their accessories (laptops, PDAs, headphones etc.). In 1998 Ericsson, Nokia, IBM, Toshiba and Intel formed the Bluetooth Special Interest Group (SIG). Their intention was to produce a global and open specification to make it possible for everyone to develop Bluetooth products and promote it on the market. Up until now, over 2200 companies have joined the SIG and an ever increasing number of Bluetooth certified products have reached the market. Certified products are marked with the Bluetooth logo, shown in figure 2.1. The Bluetooth specifications, currently in version 1.1, define a radio frequency wireless communication interface and the associated set of communication protocols and usage profiles [1].

## 2  BLUETOOTH RADIO SYSTEM ARCHITECTURE

Bluetooth devices operate at 2.4GHz, in the globally available, license-free, ISM band. That is the bandwidth reserved for general use by Industrial, Scientific and Medical applications worldwide. Since this radio band is free to be used by any radio transmitter as long as it satisfies the regulations, the intensity and the nature of interference can't be predicted. Therefore, the interference immunity is very important issue for Bluetooth. Generally [8], interference immunity can be obtained by interference suppression or avoidance. Suppression can be obtained by coding or direct-sequence spreading, but the dynamic range of interfering signals in ad hoc networks can be huge, so practically attained coding and processing gains are usually inadequate. Avoidance in frequency is more practical. Since ISM band provides about 80MHz of bandwidth and all radio systems are band limited, there is a high probability that a part of the spectrum can be found without a strong interference.

Considering all this, FH-CDMA (Frequency Hopping - Code Division Multiple Access) technique has been chosen to implement the multiple access schemes for the Bluetooth. It combines a number of properties, which make it the best choice for an ad hoc radio system. It fulfills the spreading requirements set in the ISM band, i.e. on average the signal can be spread over a large frequency range, but instantaneously only a small part of the bandwidth is occupied, avoiding most of potential interference. It also doesn't require neither strict time synchronization (like TDMA), nor coordinated power control (like DS-CDMA). In the 2.45GHz ISM band, a set of 79 hop carriers has been defined, at 1MHz spacing. A nominal hop dwell time is 625 us. Full-duplex communication is achieved by applying time-division duplex (TDD), and since transmission and reception take place at different time slots, they also take place at different hop carriers. A large number of pseudo-random hopping sequences have been defined, and the particular sequence is determined by the unit that controls the FH channel. That unit is usually called the *master* and it also defines timing parameters during the certain session. All other devices involved in the session, the *slaves*, have to adjust their spreading sequences and clocks to the master's.

Bluetooth uses Gaussian-shaped frequency shift keying (GFSK) modulation with a nominal modulation index of k=0.3. This binary modulation was chosen for its robustness, and, with the accepted bandwidth restrictions, it can provide data rates to about 1Mbps. A non coherent demodulation can be accomplished by a limiting FM discriminator. This simple modulation scheme allows the implementation of low-cost radio units, which is one of the main aims of the Bluetooth system.

An FH Bluetooth channel is associated with the piconet. As mentioned earlier, the master unit defines the piconet channel by providing the hop sequence and the hop phase. All other units participating in the piconet are slaves. However, since the Bluetooth is based on peer communications, the master/slave role is only attributed to a unit for the duration of the piconet. When the piconet is cancelled, the master and slaves roles are canceled too. In addition to defining the piconet, the master also controls the traffic on the piconet and takes care of access control. The time slots are alternatively used for master and slaves transmission. In order to prevent collisions on the channel due to multiple slave transmissions, the master applies a polling technique, for each slave-to-master slot the master decides which slave is allowed to transmit. If the master has no information to send, it still has to poll the slave explicitly with a short poll packet. This master control effectively prevents collisions between the participants in the piconet, but independent collocated piconets may interfere with one another when they occasionally use the same hop carrier. This can happen because units don't check for a clear carrier (no listen-before-talk). If the collision occurs, data are retransmitted at the next transmission opportunity. Due to the short dwell time, collision avoidance schemes are less appropriate for FH system.

## 3  BLUETOOTH PROTOCOL STACK

The Bluetooth specifications define not only a radio system but cover the underlying structure. The Core Specification contains a software protocol stack similar to the more familiar Open Systems Interconnect (OSI) standard reference model for communication protocol stacks. It permits applications to discover devices, the services they offer and permission to use these services. The stack is a sequence of layers with features crossing single or multiple layered boundaries.

If we ascend the stack, we first come across the fundamental component, the radio. The radio modulates and demodulates data for transmitting and receiving over the air. The operating band of the radio is divided into 1 MHz spaced channels with a chosen modulation scheme of Gaussian Frequency Shift Keying (GFSK). Each channel is specified to signal at 1mega symbols per second, equivalent to 1 Mb/s. The radio are the baseband and Link Controller, they are responsible for controlling the physical links via the radio, assembling the packets and controlling the frequency hopping.

Progressing through the layers, the Link Manager (LM) controls and configures links to other devices. The Host Controller Interface (HCI) is above the LM layer and is probably one of the most important layers to consider as a designer. It handles communication between host and the module. The standard defines the HCI command packets that the host uses to control the module, the event packets used by the host to inform lower protocol layers of changes, the data packets for voice and data traffic between host and module and the transport layer used by the HCI packets. The transport layer can be USB (H2), RS232 (H3), UART (4) or a robust proprietary standard such as BCSP (BlueCore Serial Protocol).

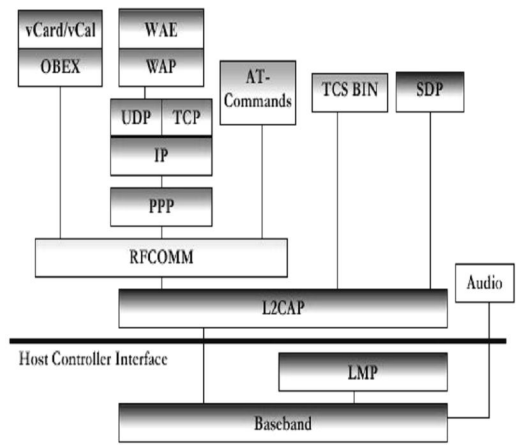

**Figure 1. Bluetooth Protocol Stack**

The Logical Link Control and Adaptation (L2CAP) is a multiplexer, adapting data from higher layers and converting between different packet sizes. The next 4 layers could be loosely grouped as communication interfaces. These are [10] RFCOMM (Radio Frequency Communication port) which provides an RS232 like serial interface. Wireless Application Protocol (WAP) and Object Exchange (OBEX) are responsible for providing interfaces to other Communications Protocols. The final member of this rough grouping is the Telephony Control protocol Specification (TCS) providing telephony services. Service Discovery Protocol (SDP) lets devices discover the services available on another Bluetooth device.

## 4 BLUETOOTH NETWORK MODULE

The Bluetooth technology operates in the Unlicensed 2.4 GHz Industrial, Scientific and Medical (ISM) band and uses a frequency hopping Time Division Duplex (TDD) scheme for transmission. Bluetooth devices share 79 [3],channels 1 MHz bandwidth within the ISM band. On the channel each packets is transmitted [1], at a different frequency. Bluetooth devices exist in small ad-hoc network configuration with the ability to operate as either master or the slave; the specification also allows a mechanism for master and slave to switch their roles. The configurations can be single point, which is the simplest configuration with one master and one slave. Multipoint, called a Piconet, based on up to 7 slaves clustered around a single Master. And a third type called a Scatternet, this is a group of Piconets effectively hubbed via a single Bluetooth device acting as a master in one Piconet and a slave in the other Piconet. The Scatternet permits either larger coverage areas or number of devices than a single Piconet can offer. Figure 2 outlines the different master and slave topologies permitted for [10] networks in the standard.

The role of the master is to control the available bandwidth between the slaves, it calculates and allocates how often to communicate with each slave and locks them into the appropriate frequency hopping sequence. The specification describes an algorithm that calculates the hop sequence, the seed being based on the master's device address and clock. In addition to hop sequence control, the master is responsible for transmit control by dividing the network into a series of time slots amongst the net members, as part of a Time Division Multiplexing (TDM) scheme. These time slots can consist of data and potentially additional voice traffic i.e. you will always need a data channel before you can add a voice channel. The time slot is defined as 625 μs and all packet traffic is allocated 1, 3 or 5 slots, grouped together in transmit and receive pairs. Prior to connection some operations such as inquiry, paging and scanning operations may sometimes occur on half slots.

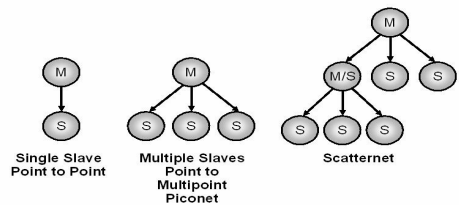

**Figure 2. Bluetooth Point to point, Piconet and Scatternet**

Bluetooth carries communication traffic over two types of air interface links defined as Asynchronous Connection Less (ACL) or Synchronous Connection Oriented (SCO). During a connection the links carry voice and data traffic in the time slots and are categorized as either time critical, as used for voice and audio, or high speed non-time critical data with a mechanism for acknowledgement and re-transmission. The first link established between master and slave is the ACL link and carries high speed data that is insensitive to time. It is packet switched, as the data is sporadic in nature, asynchronous, contains asymmetric and symmetric services and uses a polling access scheme. A master may be permitted to have a number of ACL links up to the maximum number of slaves permitted by the specification but only one link is allowed between any two devices [2].

Once an ACL has been established a SCO link can be created on top of the ACL link. The SCO link is circuit switched, it has symmetric synchronous services and has slot reservation at fixed intervals, making it suitable for time critical data such as voice. The specification restricts the number of SCO links that a master can support to three. Summarizing the two types of links:

ACL
- Packet constructed of a 72 bit access code, a 54 bit packet header, a 16 bit CRC and Payload data
- Largest data packet is DH5 giving 723.2 Kb/s as highest data rate in one direction.
- Non time critical data
- Asynchronous
- Packet switched
- Polling access

SCO
- Same Access Code and header as ACL packets
- ARQ (Automatic Repeat Request) and SEQ (Sequence) flags redundant since flow control and re-transmissions do not apply
- Cyclic Redundancy Code (CRC) field is absent
- Payload fixed at 30 bytes, with source data of 10, 20 or 30 bytes
- Circuit switched
- Symmetric synchronous services
- Slot reservation at fixed intervals

A special case exists that mixes SCO and ACL packets. Known as the Data Voice (DV) packet it carries data and voice on regular intervals like the SCO. The voice data has no flow control or CRC as per SCO packets, whereas the data part of the DV packet has flow control, re-transmission of the data part is permitted and the data part is CRC protected[3].

## 5 ROUTING IN AD HOC NETWORKS

Routing is more complex in ad hoc networks because of their dynamic nature. The supposition that nodes participate in the routing implies that network management information must be distributed among the nodes. In infrastructure-based networks this is done by protocols like RIP and BGP. However those protocols are not robust when deployed in a highly dynamic environment. In ad hoc networks, nodes might move very fast causing the entire network topology to change several times before the traditional protocols have computed new routing tables. Another major problem is the wireless environment itself. In a WLAN bandwidth is limited, collisions occur often and the link quality is seldom perfect. The wireless nodes themselves can vary from laptops with powerful processors to PDAs with very little processing power thus algorithms should not rely on the availability of high computing power. Most wireless tools run on batteries so energy also becomes a constraint as transmitting packets demands a lot of power [4].

## 6 REACTIVE ROUTING PROTOCOLS

On-demand routing protocols were designed to reduce the overheads in proactive protocols by maintaining information for active routes only. This means that routes are determined and maintained for nodes that require sending data to a particular destination.

Route discovery usually occurs by flooding a route request packets through the network. When a node with a route to the destination (or the destination itself) is reached a route reply is sent back to the source node using link reversal if the route request has traveled through bi-directional links or by piggy-backing the route in a route reply packet via flooding.

Reactive protocols can be classified into two categories source routing and hop-by-hop routing. In Source routed on-demand protocols each data packets carry the complete source to destination address. Therefore, each intermediate node forwards these packets according to the information kept in the header of each packet. In hop-by-hop routing (also known as point-to-point routing) [5], each data packet only carries the destination address and the next hop address. Therefore, each intermediate node in the path to the destination uses its routing table to forward each data packet towards the destination. The advantage of this strategy is that routes are adaptable to the dynamically changing environment of MANETs, since each node can update its routing table when they receiver newer topology information and hence forward the data packets over newer and better routes.

## 7 EXISTING PROBLEM

In the Bluetooth each and every node in the network maintains routing information to every other node in the network. Routes information is generally kept in the routing tables and is periodically updated. These routing protocols maintain different number of tables. They are not suitable for larger networks, as they need to maintain node entries for each and every node in the routing table. In the Table driven routing protocols such as [16], each node actively maintains a routing table no matter it has a message to send or not. The main drawback of the table driven routing table is the overhead to maintain the routing table at each node. Also the size of the routing table is proportional to the size of the network, the table driven protocol may require more memory than a small Bluetooth device can accommodate. Moreover if any routing table entries are not correct either through a miss configuration or through learned routes that do not match then, that resulting topology is a cyclic graph which leads Bluetooth device numerous problems such as routing loops. There are two types of forms routing loop that occur Transient and Persistent. Transient loops occur as part of the normal Operation of the routing protocol due to the different delays in [13] the propagation of information to different parts of the network. Persistent loops occur when the packets are forwarded in the optimal direction according to the information in the local routing table. If the routing table entries on all the routers are correct, the packet takes the optimal path from the source to the destination. If any routing table entries are not correct misconfiguration then these type of routing loops can occur in (figure 3). according to the routing table on Router 1, the optimal route to Network 10 is through Router 2. According to the routing table on Router 2, the optimal route to Network 10 is through Router 3. According to the routing table on Router 3, the

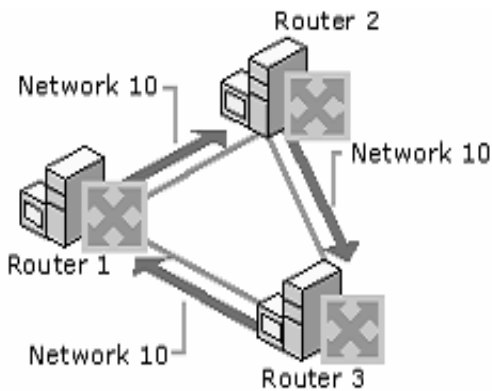

**Figure 3. Routing Loop**

Optimal route to Network 10 is through Router1. Then packet loss increase and nodes often change their location within network. So, some stale routes are generated in the routing table which leads to unnecessary [14], routing overhead. Another drawback it is not supported increased mobility and change topology.

## 8 PROPOSED APPROACH

In the Bluetooth we are using a reactive routing protocol AODV These protocols are also called reactive protocols since they don't maintain routing information or routing activity at the network nodes if there is no communication. If a node wants to send a packet to another node then this protocol searches for the route in on-demand manner and establishes the connection in order to transmit and receive the packet because of its reactive nature AODV can handle highly dynamic behavior of Ad-hoc networks used for both unicasts and multicasts Thus reactive protocols are most suited for networks with high node mobility or where the nodes transmit data infrequently It uses bandwidth efficiently is responsive to changes [11] in topology is scalable and ensures loop free routing.

## 9 AD-HOC ON-DEMAND DISTANCE VECTOR (AODV) PROTOCOL

AODV is a very simple, efficient, and effective routing protocol for Mobile Ad-hoc Networks which do not have fixed topology. This algorithm was efficient use by the limited bandwidth that is available in the media that are used for wireless communications [7].

### 9.1 Working of AODV

Each node in the network maintains a routing table with the routing information entries to its Neighboring nodes, and two separate counters: a node sequence number and a broadcast-id. When a node (source node 'S') to communicate with another destination node 'D') it increment It broadcast-id and initiates path discovery by broadcasting a route request packet RREQ to it neighbors. The RREQ contains the following field.

– source- addr

– source- sequence # -to maintain freshness info about the route to the source.

– dest- addr

– dest- sequence # - specifies how fresh a route to the destination must be before it is accepted by the source.

– hop- cnt

The (source-addr, broadcase-id) pair is used to identify the RREQ uniquely. Then the dynamic table entry establishment begins at all the nodes in the network that are on the path from S to D. As RREQ travels from node to node, it automatically sets up the reverse path from all these nodes back to the source. Each node that receives this packet records the address of the node from which it was received. This is called Reverse Path Setup. The nodes maintain this info for enough time for the RREQ to traverse the network and produce a reply to the sender and time depends on network size. If an intermediate node has a route entry for the desired destination in its routing table, it compares the destination sequence number in its routing table with that in the RREQ. If the destination sequence number in its routing table is less than that in the RREQ, it rebroadcasts the RREQ to its neighbors. Otherwise, it unicasts a route reply packet to its neighbor from which it was received the RREQ if the same[9], request was not processed previously (this is identified using broadcase-id and source-addr).Once the RREP is generated it travels back to the source, based on the reverse path that it has set in until traveled to this node. As the RREP travels back to source each node along this path sets forward pointer to the node from where it is receiving the RREP and records the latest destination sequence number to the request destination . This is called Forward Path Setup. If an

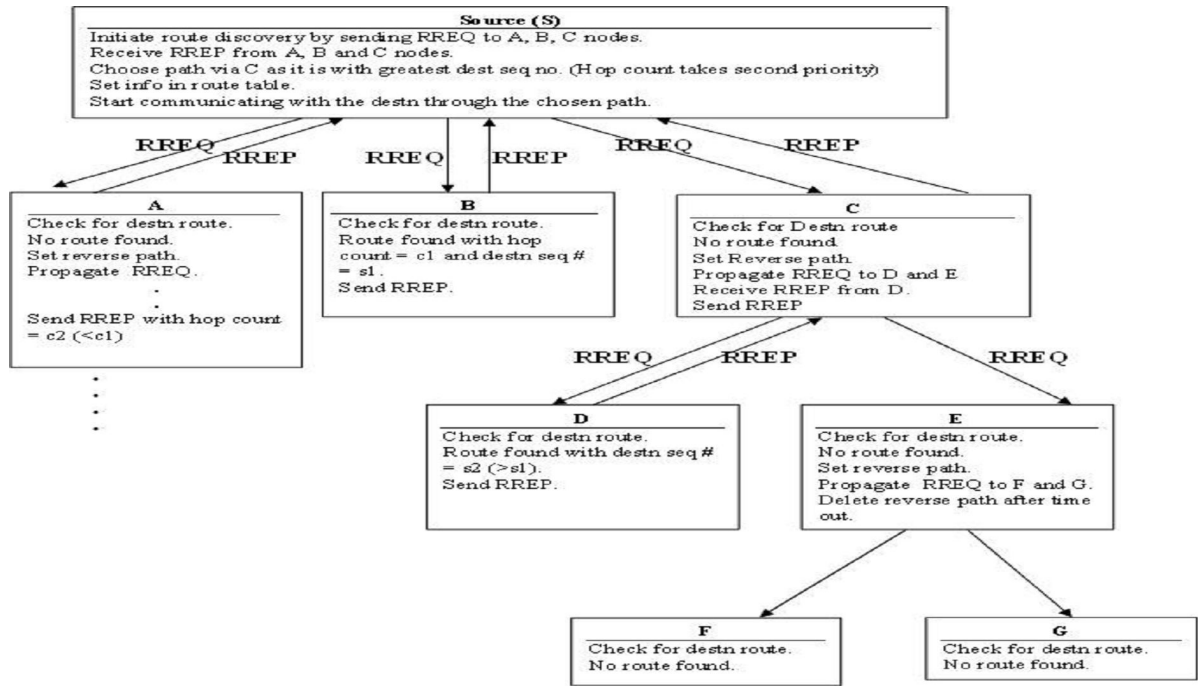

**Figure 4. Route finding process in AODV Routing Protocol**

Intermediate node receives another RREP after propagating first RREP towards source it checks for destination sequence number of new RREP. The intermediate node updates routing information and propagates new RREP only.

– If the Destination sequence number is greater, OR

– If the new sequence number is same and hop count is small,

Step by step explanation of below figure

1. Source 'S' has to send data to destination.
2. S sends RREQ to its neighbors A, B, C.
3. B finds the path in its routing table (destn seq-number s1 and hop count c1) and sends RREP to S.
4. C sets up reverse path.
5. C forwards RREQ to its neighbors D and E.
6. E sets up reverse path.
7. E forwards RREQ to its neighbors F and G.
8. E deletes the reverse path after a time out period as it does not receive any RREPs from F and G.
9. D finds the path (dest seq-number s2 which is greater than s1 and hop count c1) in routing table and sends RREP to C.
10. C receives RREP from D and sets up forward path and forwards RREP to S.
11. A sets reverse path forwards RREQ to its neighbors receives RREP (path of hop count c2 which is greater than c1) sets forward path and forwards this RREP to S.
12. S receives a path info from C ( with destn seq-number s2 and hop count c1 ), another path info from B ( with destn seq-number s1 and hop count c1), and another path info from A( destn seq-number x which is less than s1 and s2 and hop count c2 which is less than c1).
13. S chooses path info from C (which was originated from D), giving first priority to the path greatest destination sequence number and then second priority to the path with smallest hop count. Though path given by A is of smallest hop count, it is ignored because the destination sequence number is greater than the path from C.

### 9. 2 Route Table Management

In each routing table entry, the address of active neighbors which packets for the given destinations are received is also maintained .A neighbor is considered active in other words for that destination if it originates or relays at least one packets for that destination within the most recent active timeout period [15] .This information is maintained, so that all active source nodes can be notified when link along a path to the destination breaks. A route entry is considered active if it is in use by any active neighbors. The path from a source to destination, which followed by packets along active route entries, is called an active path. Note that AODV, all routes in the route table are tagged with the destinations sequence number , which guarantee that no routing loop can form even under extreme conditions of out of order packet delivery and high node mobility .

Each mobile node in the network maintains a route table entry for each destination of interest in its route table. Each entry contains the following info:

– Destination

– Next hop

– Number of hops

– Destination sequence number

– Active neighbors for this route

– Expiration time for the route table entry

## 10 PERFORMANCE METRICS

There are three important performance metrics are evaluated.

**-- Average delay of data packet:** This includes all possible delay examples for route discovery latency, retransmission delay at the MAC, and propagation and transfer time.

**-- Packet Loss:** This includes the number of packet send and number of packet received.

**-- Throughput:** Number of packet received in a particular interval of time.

## 11 EXPERIMENTAL RESULTS

The simulation results were generated with NS-2.0 version ns-allinone-2.30 (NS-2, 2004) and BlueHoc Bluetooth simulator (Tan, 2002). The simulation environment has been established under a [12] Red Hat (rel.11) Fedora Linux 2 operating system with GCC 2.95.5 (Fedora) compiler installed. The hardware included Intel Pentium 4 processor, with 512 MB DDRAM, and 40 GB Hard Disk. Next sections explain the details of each experiment. These three parameters we are also using to measure of the propose protocol.

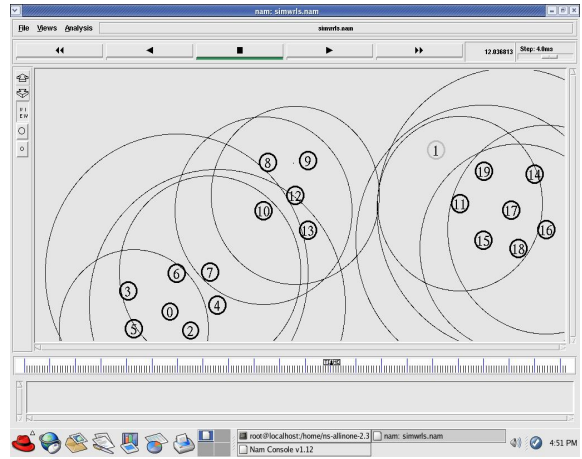

**Figure 5. Create Scatternet and Change the Topology**

We created a graphical animation of the scenario in NAM (Network Animator) from the NS-2 output. There are three Piconet create in ns-2 .In piconet 1st their are 7 salve and 1 is master (Node no.1) same as piconet 2nd it contain a 5 node , node number 12 is master and rest of slave , and same as piconet 3rd its also contain a 7 node , node number 17 is master and rest of slave. In this figure 5 we are change the Topology , means in our piconet 1st node number 1 is rapidly traverse from piconet 1st to piconet 3rd and this node are both piconet 2nd and piconet 3rd act as salve and communicate. These result of the Simulation of the AODV routing protocol and measured by using the following parameters given in the below.

-- Delay
-- Throughput
-- Packet Loss

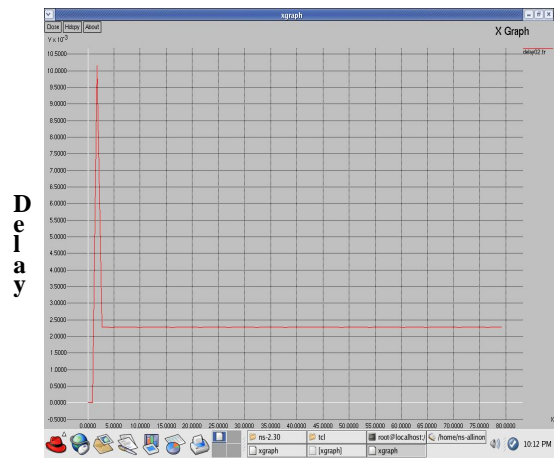

**Transmission Time (ms)**
**Figure 6. Delay X Graph**

At time 0.5 nodes is transmitting of packet so the all available path in source to destination in our scatternet are used by this node as result delay of this node is low in figure 6.

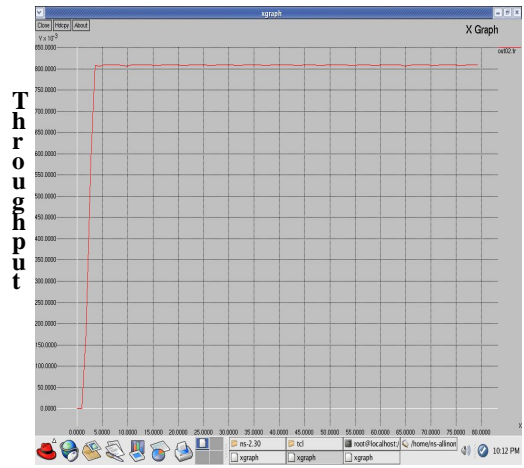

**Transmission Time (ms)**
**Figure 7. Throughput X Graph**

In this figure 7 at time 0.5 when nodes is started transmission of packet from source to destination in the available path of the scatternet then throughput is very high. The packet loss is very low when node is started transmission of packet from source to destination in the available path of the Scatternet in figure 8.

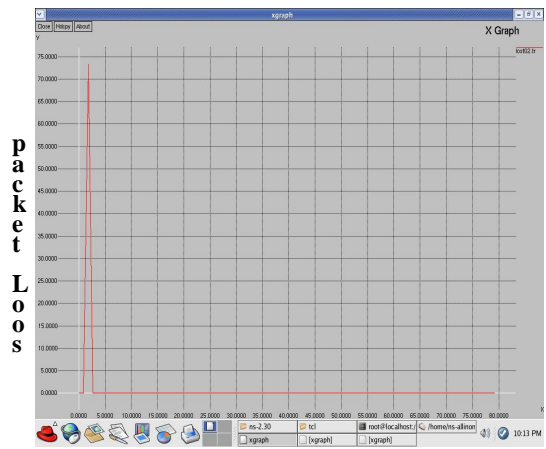

**Transmission Time (ms)**
**Figure 8. Packet Loss X Graph**

## 11.1 Simulation parameters used

| Parameters | values |
|---|---|
| Channel | Wireless |
| Propagation Model | Two Ray Ground |
| Mac | Mac/802_11 |
| Antenna | Omni Directional |
| No of Mobile Node | 20 |
| Area(x,y, Dimension ) | 500*400 |
| Routing Protocol | AODV |
| Queue Length | 50 |
| Basic Rate | 5MB |
| Data Rate | 10MB |

**Table 1. Simulation parameters for scatternet network**

## 12 CONCLUSION

The conclusion of the project depicts that there are some drawbacks with the existing Bluetooth Scatternet. So we have proposed a Reactive Routing Protocol technique to achieve the free routing Loop in the scatternet, also we enhanced the mobility and this lead to the change to the topology.

During the transmission we found that the packet loss is low at the same time the throughput is high.

The results showed the graphical animation of the overall scenario in NAM (Network Animator) and the X-Graph. Bluetooth Scatternet Communication might become important in the time to come , because communication devices with reduced physical dimensions and low cost could have a significant growth & impact in the fields of medical, military, and corporate business markets in the near future.

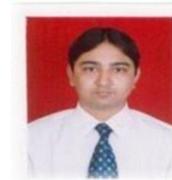
**Yusuf Perwej** received the M .Tech (IT) degree in Computer Science and Engineering from the C - DAC ( Centre for Development of Advanced Computing ) Govt.of India Research & Development center,Noida (India) in 2007, an MCA degree in computer application from the University of GGSIP, Delhi (India) in 2005 .He is currently the department of Research and development member and Senior Lecturer in the department of Computer Science & Information System at University of Jazan , KSA. He has authored a number of different journal and paper. His research interests include wireless and mobile computing system, soft computing, ad hoc network, cryptography, artificial neural network, artificial intelligence, robotics, Bluetooth, routing protocol for ad hoc network, as well as Software Engineering.

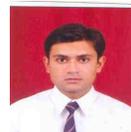
**Dr. Kashiful Haq** received the Doctor of Philosophy (Ph.D.) in Mathematics (Operations Research) from the University of CCS, Meerut (India) in 2008, an M.A degree in Mathematics from the University of CCS, Meerut (India) in 2004 .He is currently the department of Research and development member and Assistant Professor in the IT department at the College of Applied Sciences, Ministry of Higher Education, Ibri, Sultanate of Oman .He has authored a number of different journal and paper. His research interests include Discrete Mathematics ,Graph Theory, Software Engineering , Mathematics, wireless and mobile computing system, ad hoc network, Operations Research, MIS, cryptography, routing protocol for ad hoc network.

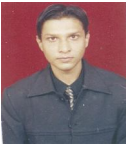 **Uruj Jaleel** received the M .Tech degree in Computer Science and Engineering from the Shobhit University, Meerut , (India) in 2009, an MCA degree in computer application from the Meerut Institute of Engineering &Technology, Meerut, affiliated from the UPTU,UP(India) in 2004 .He is currently the department of Research and development member and Senior Lecturer in the department of Computer Science & Information System at University of Jazan ,KSA.He has authored a number of different journal and paper. His research interests include wireless and mobile computing system, ad hoc network, cryptography, E-Commerce, Image processing, routing protocol for ad hoc network, Software Engineering.